\documentstyle[12pt,twoside]{article}
\include {def}
\newcommand{\noi}{\noindent}
\begin{document}

\thispagestyle{empty}
\vspace*{30pt}

\noi
{\large Interplay of  Correlation,  Randomness and Dimensionality Effects
  in  Weakly-Coupled Half-Filled Random Hubbard Chains}

\vspace{1cm}
\noi
JUN-ICHIRO KISHINE  AND KENJI YONEMITSU\\
Institute for Molecular Science, Okazaki 444-8585, Japan

\vspace{1cm}

\noi
We study interplay of electronic correlation, randomness and dimensionality effects in half-filled 
random Hubbard chains weakly coupled via an interchain one-particle hopping. Based on the 
two-loop renormalization-group approach, phase diagrams  are given 
in terms of temperature  vs. strengths of the intrachain electron-electron umklapp scattering, 
the random scattering and the interchain one-particle hopping.

\vspace{1cm}
\noi
\vspace{1cm}

\baselineskip17.8pt

\noi
{\bf INTRODUCTION}

\noi
Randomness effects in interacting electron systems have provoked a great deal of controversy over the past two decades. 
In the one-dimensional case away from half filling, Giamarchi and Schulz$^{[1]}$ discussed strong interference between 
correlation-induced quantum fluctuations and random scattering processes, by applying the renormalization-group (RG) approach 
to the bosonized Hamiltonian obtained by the replica trick. Recently, Fujimoto and Kawakami$^{[2]}$ discussed the case of half 
filling, where  competition between the correlation-induced Mott transition and the  randomness-induced Anderson 
localization arises.
In quasi-one-dimensional systems, three dimensionality induced by an interchain coupling plays an important role$^{[3]}$.
 In half-filled Hubbard chains weakly coupled via an interchain one-particle hopping, $t_{\perp}$, without randomness, 
there occur the interchain one-particle propagation through the $t_{\perp}$-
process and the propagation of the 1D antiferromagnetic (1DAF)  power-law
correlation through the interchain particle-hole 
exchange (ICEX) processes$^{[3],[4]}$. The former process drives a crossover to the Fermi liquid (FL) 
regime, while the latter process causes a transition to an antiferromagnetic (AF) phase   from an incoherent metal (ICM)
 phase$^{[4]}$. In this paper, we study the randomness effects on the competition between them
 in weakly-coupled half-filled random Hubbard chains. 
We apply the two-loop RG approach to the  effective action obtained by the replica trick. Instead of the bosonization 
technique$^{[1],[2]}$ which takes account of only the  collective degrees of freedom, we here use the effective action with
 Grassmann representation to treat
both of the interchain one-particle (individual) and two-particle (collective) processes.

\vspace{1cm}
\noi
{\bf EFFECTIVE ACTION}

\noi
We consider an   array of the half-filled random Hubbard chains. In Figs.1(a)-1(e), we show all 
the fundamental processes included here. The intrachain   one-particle propagation to the right- and left-directions 
[Figs.1(a-1) and (a-2), respectively] is treated by linearizing the one-particle dispersion at the Fermi 
points. Scattering of electrons by the weak random potential at a spatial position, $x$, is taken into account through 
a real field, $\eta(x)$, corresponding to random forward scattering [Fig.1(b-1)] and 
complex fields, $\xi(x)$ and $\xi^\ast(x)$, corresponding to random backward scattering [Fig.1(b-2)]. We assume the random 
potential to be governed by Gaussian distributions,
\begin{eqnarray}
P_\eta &\propto&\exp\left[-D_\eta\int d{ x}\eta(x)^2\right],\\
P_\xi &\propto&\exp\left[-D_\xi\int d{  x}\xi(x)\xi^\ast(x)\right],
\end{eqnarray}
where  $D_\eta=(\pi N_F \tau_\eta)^{-1}$ and  $D_\xi=(\pi N_F \tau_\xi)^{-1}$
with $\tau_{\eta,\xi}$  and $N_F$  being the scattering mean free times and the noninteracting one-particle density of states, 
respectively. At half filling, the intrachain interaction generates the normal [Figs.1(c-1,2)] and umklapp [Fig.1(c-3)]
 scattering processes with dimensionless strengths, $g_1$, $g_2$  and $g_3$, for 
the backward, forward and umklapp scattering, respectively. Initial values of the 
intrachain scattering strengths are related to the on-site Coulomb repulsion, $U$ , by
$g_{1;0}=g_{2;0}=g_{3;0}=U/\pi v_F$ at half filling with $v_F$
  being the Fermi velocity. We here consider only the case of
$g_{1;0}-2g_{2;0}<\mid\! g_{3;0}\!\mid$, where
  the most dominant 1D power-law correlation is an 
antiferromagnetic one. Accordingly, the most dominant ICEX  process is in the AF channel. 
Figures 1(d-1) and (d-2)  show the interchain one-particle hopping ($t_\perp$) processes between the nearest-neighbor
 chains for the right- and left-moving electrons, respectively. 
Multi-$t_\perp$- and $g_i\,(i=1,2,3)$-scattering processes generate dynamically 
the interchain AF interactions  during the renormalization. 
Their strengths in the normal  and umklapp
channels are denoted by $J$~[Fig.~(e-1)] and $K$~[Fig.~(e-2)], respectively. 
        
\vspace{7cm}
\noi{\bf FIGURE 1} 
Fundamental processes considered here. The solid and broken lines represent the propagators for the 
right- and left-moving electrons, respectively.  The wavy lines 
represent the intrachain two-particle scattering~($g_1$, $g_2$ and $g_3$). The zigzag lines represent the interchain one-particle
 hopping, $t_\perp$, from the $i$-th chain to the nearest-neighbor
 chain. The white and black squares represent the interchain  AF  interactions, $J$ and $K$, in
 the normal and umklapp channels, respectively. $i$ and $j$ denote different chain indices.
\vspace{1cm}
  
 We consider   quenched randomness, where   the 
free energy is averaged over  random potentials  by  the replica trick,   $\ln Z=\lim_{N\to 0}(Z^N-1)/N$, for
the partition function, $Z$. Here $N$  identical replicas are introduced with   indices, 
$\alpha=1,2,\cdots,N$. We take 
an average $\langle Z^N \rangle_{\rm random}$ for integer $N$, continue the result analytically to real $N$, and finally take 
the  $N\to 0$ limit.
 
\vspace{1cm}
\noi
{\bf RENORMALIZATION-GROUP EQUATIONS}

\noi
Based on the bandwidth cutoff regularization 
scheme, we parameterize the temperature as $T=E_0e^{-l}$
 with the scaling parameter $l$.
As $l$ goes from zero to infinity, we move from high-temperature scales, where the system 
is regarded as collection of isolated 1D chains, to low-temperature scales, where the interchain 
coupling is substantial. Performing a scale transformation at the two-loop level and re-expressing the 
renormalized action at   $l+dl$   in terms of  the renormalized coupling strengths,
 we obtain RG equations,
\begin{eqnarray}
{d\tilde g_{1} / dl}&=& w_{1}- 2\theta\tilde g_{1}-\tilde D_{\xi} , \label{eqn:g1}\\
{d\tilde g_{2}/ dl} &=&w_{2}- 2\theta\tilde g_{2}-\tilde D_{\eta},\\
{dg_{3}/ dl}&=&w_{3} -2\theta g_{3} ,\\
{d\tilde D_{\eta}/ dl}&=&w_{\eta}+ (1  -2\theta )\tilde D_{\eta},\\
{d\tilde D_{\xi}/ dl}&=&w_{\xi}+ (1-2\theta )\tilde D_{\xi}, \\
{d\ln t_{\perp}/ dl} &=&1-\theta, \\
{dJ / dl}  &=&-  ({\tilde g_{2}} ^2+4{g_{3}}^2 ) 
\tilde t_{\perp}^2/2 
+ \tilde g_{2} J/2 +2g_{3}  K -  J^2/4-K^2  ,  \\
{d K/ dl} 
&=&-2 
{\tilde g_{2}}{g_{3}}\tilde t_{\perp}^2 
+2 (\tilde g_{2}  K +g_{3} J   )-JK ,\label{eqn:K}
\end{eqnarray} 
where 
$\tilde D_{\xi}={D_{\xi}\Lambda /2v_F}$,
$\tilde D_{\eta}={D_{\eta}\Lambda /2v_F}$,
$\tilde g_{1}=g_1-\tilde D_{\xi}$, $\tilde g_{2}=g_2-\tilde D_{\eta}$ and $\tilde t_{\perp}  = t_{\perp}/ E_0 $.
$\Lambda$ is the length scale which characterizes  inelastic scattering processes between   different replicas.
The two-loop vertex correction diagrams for $\tilde g_{1}$, $\tilde g_{2}$, $g_3$, $\tilde D_{\eta}$ and $\tilde D_{\xi}$ give
$w_{1}$, $w_{2}$, $w_{3}$, $w_{\eta}$ and $w_{\xi}$, respectively, and the two-loop self-energy diagrams give $\theta$.
The contribution of the diagrams containing a  loop connected to outer lines via the random scattering 
 as shown in Fig.~2 is proportional to the number of replicas, $N$, and   vanishes in the replica limit, $N\to 0$.     

\vspace{3cm}
\noi{\bf FIGURE 2} 
A  diagram which vanishes in the replica limit, $N\to 0$.   
         Note that the replica index of the internal loop, $\gamma$, runs from $1$ to $N$.
\vspace{1cm}

\noi{\bf PHASE DIAGRAMS}

 \noi
Based on the RG flows obtained through solutions of Eqs. (\ref{eqn:g1}) -- (\ref{eqn:K}), 
we introduce the three characteristic  scales, $l_{\rm loc}$, $l_{\rm FL}$
and $l_{\rm AF}$:
\begin{eqnarray}
\tilde D_{\xi}=1&& \,\,\, {\rm at}\,\,l=l_{\rm loc},\\
t_{\perp}/E_0=1&& \,\,\, {\rm at}\,\,l=l_{\rm FL},\\
K=-\infty&& \,\,\, {\rm at}\,\,l=l_{\rm AF}.
\end{eqnarray}
$l_{\rm loc}$ is the scale which characterizes the crossover to the Anderson 
localization phase.
 We call this phase a \lq\lq one-dimensional Anderson localization (1DAL)
 phase,\rq\rq \,
 since   the one-dimensional 
scaling procedure works during the crossover.  $l_{\rm FL}$
is the  scale which characterizes the crossover to the three-dimensional
 FL  phase with randomness.  $l_{\rm AF}$ 
is the scale at which there occurs the phase transition to 
the  AF  phase from the  ICM  phase. We solve the
 coupled RG equations, (\ref{eqn:g1}) -- (\ref{eqn:K}), and check which of Eqs. (11) -- (13) is satisfied 
at the highest energy scales, depending on the initial values of the coupling strengths, and obtain phase
 diagrams which are classified into the following three categories. 

\vspace{7.5cm}
\noi{\bf FIGURE 3} Phase diagrams of the system.
{\bf ICM}=incoherent metal phase.  
{\bf AF}=antiferromagnetic phase. 
{\bf 1DAL}=one-dimensional Anderson localization phase.   
{\bf FL}=Fermi liquid phase.
\vspace{1cm}

\noi
{\bf I.} \underline{Competition between the AF phase and the 1DAL phase [Fig.3(a)]}

For $U=0.3\pi v_F$,
$\tilde D_{\eta;0}=0.01$ 
 and $t_{\perp;0}=0.01E_0$, 
we obtain   Fig.3(a) 
on the plane  of the initial random backward scattering strength, $\tilde D_{\xi;0}$,
 and the temperature. Owing to the strong umklapp scattering   and the weak interchain one-particle hopping, 
the low temperature phases are determined by the competition between the AF coherence and the 1DAL. We 
see that the AF transition temperature, $T_{\rm N}$, increases with  $\tilde D_{\xi;0}$, since during the renormalization
 process growth of  $\tilde D_{\xi}$ suppresses $\tilde g_1$ and enhances $g_3$. 
Growth of the umklapp scattering plays an essential role for the AF transition from the ICM phase$^{[4]}$.

\noi
{\bf II.} \underline{Competition between the 1DAL phase and the FL phase [Fig.3(b)]}

 For $U=0.05\pi v_F$, $\tilde D_{\eta;0}=0.01$  and $\tilde D_{\xi;0}=0.2$,
 we obtain   Fig.3(b)  
on the plane of the initial  interchain one-particle hopping integral, $t_{\perp;0}$, 
and the temperature. Owing to the weak umklapp scattering
 and the strong random backward scattering, the low temperature phases are determined by
 the competition between the 1DAL and the interchain one-particle coherence.

\noi
{\bf III.} \underline{Competition between the AF phase and the  FL phase [Fig.3(c)]}

For $U=0.5\pi v_F$,  $\tilde D_{\eta;0}=0.08$ and $\tilde D_{\xi;0}=0.08$, we obtain   Fig.3(c)  on the plane
 of   $t_{\perp;0}$ 
 and the temperature. Owing  to the strong umklapp scattering and the weak random backward scattering, the low temperature phases are determined by
 the competition between the AF coherence and the interchain one-particle coherence.

 Finally, it should be noted that even inside the AF   and   FL phases in   Figs.3(a)-(c),  the randomness effects would
 work. This issue is beyond the scope of the present  RG approach.

\vspace{0.8cm}
\noi
{\bf Acknowledgments}

\noi
This work was supported by a Grant-in-Aid for Encouragement of Young Scientists from the Ministry of Education, Science, Sports and Culture, Japan.
\vspace{0.8cm}

\noi
{\bf
References}

\noi [1]T.Giamarchi and H.J.Schulz, {\it Phys. Rev.} {\bf \it B}{\bf 37}, 325 (1988).

\noi [2]S.Fujimoto and N.Kawakami, {\it Phys. Rev.} {\bf \it B}{\bf 54}, 11018 (1996).

\noi [3]C.Bourbonnais, in {\it Strongly Interacting Fermions and High Tc}
{\it Super-conductivity}, ed. B.Doucot and J.Zin-Justin (Elsevier,  1995), p.307.

\noi [4]J.Kishine and K.Yonemitsu, {\it J. Phys. Soc. Jpn.} {\bf 67}, 2590 (1998); 
J.Kishine and K.Yonemitsu, {\it J. Phys. Soc. Jpn.} {\bf 68}, No.8 (1998).

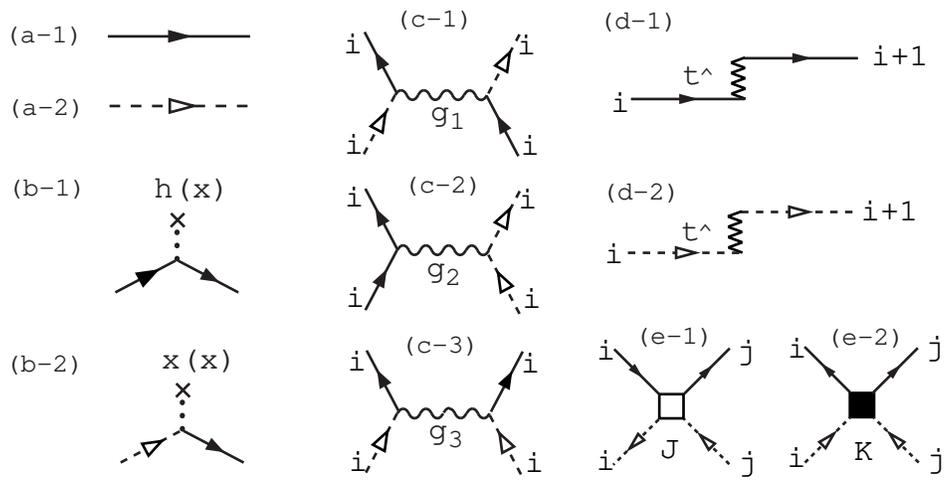
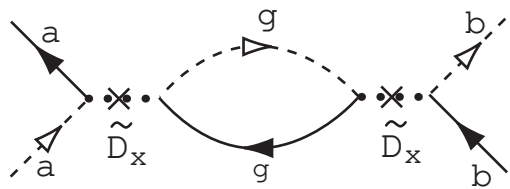
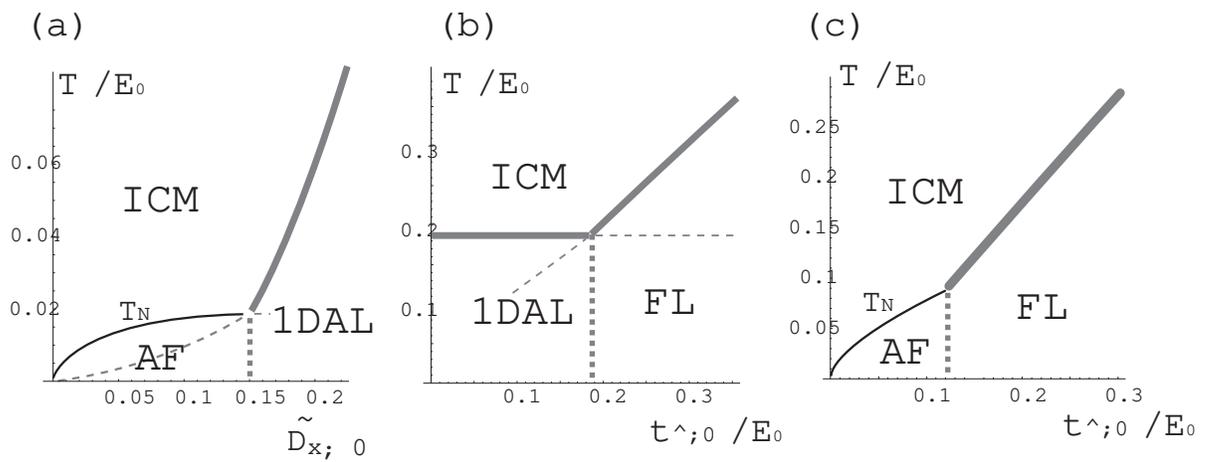
\end{document}